\documentclass[aps,pre,preprint]{revtex4-1}

\usepackage{amsmath}
\usepackage{amssymb}
\usepackage{graphicx}
\usepackage{subfig}
\usepackage{color}

\DeclareGraphicsExtensions{.pdf,.png}

\begin{document} 

\title{Properties of low--dimensional collective variables in the molecular dynamics of biopolymers}
\author{Roberto Meloni}
\affiliation{Department of Physics, Universit\`a degli Studi di Milano, and INFN, via Celoria 16, 20133 Milano, Italy}
\author{Carlo Camilloni}
\affiliation{Department of Chemistry and Institute for Advanced Study, Technische Universit\"at M\"unchen, Lichtenbergstr. 4, 85747 Garching, Germany}
\author{Guido Tiana}
\email{guido.tiana@unimi.it}
\affiliation{Center for Complexity and Biosystems and Department of Physics, Universit\`a degli Studi di Milano, and INFN, via Celoria 16, 20133 Milano, Italy}
\date{\today}
\begin{abstract}
The description of the dynamics of a complex, high--dimensional system in terms of a low--dimensional set of collective variables $Y$ can be fruitful 
if the low dimensional representation satisfies a Langevin equation with drift and diffusion coefficients which depend only on $Y$. 
We  present a computational scheme to evaluate whether a given collective variable provides a faithful low--dimensional representation of the dynamics of a high--dimensional system. The scheme is based on the framework of finite--difference Langevin--equation, similar to that used for molecular--dynamics simulations. 
This allows one to calculate the drift and diffusion coefficients in any point of the full--dimensional system. 
The width of the distribution of drift and diffusion coefficients in an ensemble of microscopic points at the same value of $Y$ indicates to which extent the dynamics of $Y$ is described by a simple Langevin equation. 
Using a simple protein model we show that collective variables often used to describe biopolymers display a non--negligible width both in the drift and in the diffusion coefficients. We also show that the associated effective force  is compatible with the equilibrium free--energy calculated from a microscopic sampling, but results in markedly different dynamical properties.
\end{abstract}

\maketitle

\section{Introduction}

Biomolecular models are usually very high-dimensional systems.
Proteins in solution, as described in classical molecular-dynamics (MD) simulations, 
are characterized by the three--dimensional positions and velocities of tens 
or hundred of thousands atoms, thus defining a $10^5$--$10^6$--dimensional phase space. 
Studying the result of a MD simulation in such a high--dimensional space is outrageously difficult, and consequently 
one usually follows the behavior of low--dimensional (often one--dimensional), 
collective variables (CVs) $Y(\mathbf{r})$ which are function of the high--dimensional 
coordinates $\mathbf{r}$ of the system \cite{Zwanzig}. 

Also in dealing with experiments, one can usually monitor only low--dimensional CVs. 
For example, the ellipticity in circular dichroism, the fluorescence intensity of tryptophanes 
or the nuclear chemical shifts in NMR spectra depend on the 
high--dimensional coordinates $\mathbf{r}$ of a system but can provide only 
a low--dimensional view of its conformational properties.

While in the design of experiments the choice of the CVs to monitor 
is determined by the technique itself, in simulations one can choose to compute 
and analyze essentially any function of the microscopic coordinates. 
The problem is  how to make such a choice. In the study of equilibrium properties 
of a system, the only feature that the CV must have is to be able 
to distinguish the relevant phases of the system, that is to be a good "order parameter". 
In the case of protein folding, for example,  it should assume different values in 
the native state, in the denatured state and, when relevant, in intermediate states. 
The root mean square deviation of the atomic positions (RMSD) with respect to those 
of the native conformation, or the fraction $q$ of native contacts usually can do that \cite{Du:1998}.

The choice is less straightforward for the description of the time--dependent, 
dynamical properties of the system \cite{Berezhkovskii:2005,Krivov:2016}. For most purposes, it is useful that the 
CV $Y$ obeys a Langevin equation of kind
\begin{equation}
\frac{dY}{dt}= D^{(1)}(Y) + \sqrt{2D^{(2)}(Y)} \cdot \eta(t),
\label{eq:red-langevin}
\end{equation}
where $\eta$ is a stochastic, Gaussian variable with moments $\overline{\eta(t)}=0$ and 
$\overline{\eta(t)\eta(t')}=\delta(t-t')$, and the bar indicates the 
average over the realisations of the stochastic variable.
In this case, $D^{(1)}(Y)$, which depends on the gradient of the equilibrium free energy $F(Y)$, can be regarded as an 
effective force acting on the CV, and thus is very rich of information 
on the system;  $D^{(2)}(Y)$ is a position--dependent coefficient which controls the diffusion 
of $Y$ within its low--dimensional space \cite{Zwanzig}. 

If Eq. (\ref{eq:red-langevin}) holds, 
one can use Arrhenius equation to estimate reaction rates, can model the dynamics 
of the system as transitions between discrete states, and can use all the armoury of tools
developed in the realm of Langevin equations \cite{Zwanzig}.
Besides the advantages in analysing and plotting {\it a posteriori} 
the relevant information, CVs satisfying 
Eq. (\ref{eq:red-langevin}) can be used on--the--fly to bias the dynamics of the 
system and thus speed-up equilibrium sampling, like in the case of umbrella sampling, metadynamics
and steered molecular dynamics \cite{Laio:2002, Zheng:2013, Abrams:2014}, 
or even to obtain efficiently dynamical properties \cite{Berezhkovskii:2005,Tiwary:2016}. 
In the following, we shall call "reaction coordinate" any CV that 
satisfies Eq. (\ref{eq:red-langevin}).

An equation formally similar to Eq. (\ref{eq:red-langevin}) can always be 
written for any CV. However, a bad choice of the CV 
results in functions $D^{(1)}$ and $D^{(2)}$ which depend not only on $Y$, but on the 
whole history of the system; in other words, in this case $Y$ undergoes a 
non--Markovian process \cite{Zwanzig}. This happens when $Y$ does not really determine the properties 
of the relevant phase space where the dynamics of the system is likely to occur, 
but, on the contrary, to a given value of $Y$ can correspond different, well--separated regions of 
the relevant microscopic phase space, in which the effective 
force $D^{(1)}$ and the effective diffusion coefficient $D^{(2)}$ are very different. 
In this case, having discarded the microscopic coordinates, only time can distinguish 
between the different phase--space regions at identical $Y$, resulting in a non--Markovian 
dynamics.

A necessary and sufficient condition for Eq. (\ref{eq:red-langevin}) to hold is 
that $Y$ describes the slowest kinetic modes of the system \cite{Zwanzig}. 
This implies that the system can visit very quickly all the phase--space accessible regions 
for any fixed value of $Y$, thus equilibrating the coordinates perpendicular to it. 
In this context, fixed means that the relative change in value of $Y$ is negligible 
with respect to that of the perpendicular coordinates.
As a consequence, $D^{(1)}$ and $D^{(2)}$ 
are determined by the average contribution of all the phase--space regions displaying 
the same value $Y$, 
the time dependence is averaged out and only the dependence on $Y$ remains. 
However, this is not a property which can be easily verified for any given function 
of the microscopic coordinates $\mathbf{r}$ of the system.

The study of the dynamics of a system by Eq. (\ref{eq:red-langevin}) requires the evaluation of $D^{(1)}$ and $D^{(2)}$ from a set of microscopic trajectories.
Given a set of stochastic trajectories $\{\mathbf{r}(t)\}$, generated with some MD algorithm from a point $\mathbf{r}(0)$ of conformational space, the effective force and diffusion coefficient defined by Eq.  (\ref{eq:red-langevin}) in that point can be obtained by definition as the first two Kramers--Moyal coefficients \cite{risken}. 
Defining $Y_t\equiv Y(\mathbf{r}(t))$, then 
\begin{align}
D^{(1)}(Y_0)=\lim_{\Delta t\to 0} \frac{1}{\Delta t}\overline{ (  Y_{\Delta t} - Y_0  ) } \nonumber\\
D^{(2)}(Y_0)=\frac{1}{2}\lim_{\Delta t\to 0} \frac{1}{\Delta t}\overline{ (  Y_{\Delta t} - Y_0  )^2 }.
\label{eq:km}
\end{align}
Physically, the limit $\Delta t\to 0$ means that  $D^{(1)}$ and $D^{(2)}$ should be a property 
of $Y$ only, independent on where the trajectories go to afterwards. 
In the case of MD simulations of biopolymers, the use of Eqs. (\ref{eq:km}) presents a serious problem, namely that any 
integrator that can be used to generate $\{\mathbf{r}(t)\}$ has a finite time step, and thus the limit $\Delta t\to 0$ cannot be evaluated. 

Several works tried to estimate the drift and diffusion coefficients of 
Eqs. (\ref{eq:km})  in the limit of small $\Delta t$, 
by correction terms \cite{Ragwitz:2001uw,Anteneodo:2009to}, by iterative procedures \cite{Kleinhans:2005vl}, or by evaluating the adjoint Fokker--Planck operator \cite{Lade:2009tk,Honisch:2011wv}. However one should stress that these works face the problem of evaluating $D^{(1)}$ and $D^{(2)}$ for generic time series of observables characterized by a low, uncontrollable, sampling rate.  In the case of MD simulations, the minimum time period can be as small as an integration time step, which is smaller than any process involved in the microscopic dynamics.

Assuming to know the reaction coordinate $Y$, an efficient way of extracting drift and diffusion coefficient from MD simulations was developed on the basis of a Bayesian approach \cite{Hummer:2005eb,Sriraman:2005jf} and then applied to protein folding \cite{Best:2006gj,Best:2010cxa}. The main result of these works is that the diffusion coefficient for variables that scale as the Euclidean distances between Cartesian coordinates  depends strongly on $Y$, while the diffusion coefficient for variables that have a filtered dependence on such distances, like contact functions, depend weakly on $Y$.

Using a maximum--likelihood principle \cite{Micheletti:2008}, the drift and diffusion coefficients could be obtained as average of molecular dynamics trajectories, and a criterion for the  choice of the sampling rate of the trajectories was introduced to minimize time correlations of noise.

The goal of the present work is rather different from those discussed above. Our primary task is that of  investigating the validity of the framework defined by Eq. (\ref{eq:red-langevin}), and in particular whether it is possible to define the drift and diffusion coefficients $D^{(1)}$ and $D^{(2)}$ as a function of  $Y$. In other words, we studied the validity of the hypothesis at the basis of refs. \cite{Hummer:2005eb,Sriraman:2005jf,Best:2006gj,Best:2010cxa,Micheletti:2008}. Another important difference is that,
since we are interested in facing the problem from a computational perspective, we did not study directly the validity of true Langevin equation (\ref{eq:red-langevin}), but its finite--differences counterpart, defined within the scheme of a standard integrator at finite time step $\Delta t$. In fact, it is the finite--difference dynamic equation what one usually calculate in MD simulations.

In next section we present an algorithm to obtain efficiently the drift and diffusion coefficient of the finite--time--step Langevin equation. Then we show to which extent $D^{(1)}$ and $D^{(2)}$ are function of $Y$ only, and not depend on the detailed microscopy coordinates. We first applied this analysis to some test models and then to some popular CVs within simple alpha--helix and beta--hairpin models. Finally, we studied to which extent the calculated coefficients are in agreement with the equilibrium free energy of the system, and the dynamics in the reduced space is in agreement with the projection of the associated microscopic dynamics.

\section{Drift and diffusion coefficients in finite--difference Langevin Equation}

\subsection{Finite--difference equation in Euler approximation}

Consider a set of $M$ trajectories $\{\mathbf{r}(t)\}$ in the microscopic conformational space, 
solution of Langevin equations (either complete or overdamped) at fixed temperature $T$, 
starting from a given point $\mathbf{r}(0)$ and integrated with a time step $\Delta t_{mic}$ making 
use of a suitable integrator. Given a function $Y(\mathbf{r})$ of the microscopic variables, we want to 
investigate whether the projection of the microscopic dynamics on $Y$ can be described in terms 
of the Euler--integrator version of overdamped Langevin equations
\begin{equation}
Y_{t+\Delta t}=Y_t+D^{(1)}(Y_t)\Delta t+\sqrt{  2 D^{(2)}(Y_t)\Delta t}\cdot\eta_t
\label{eq:euler}
\end{equation}
where $\eta_t$ is an adimensional Gaussian--distributed stochastic variable with zero average and 
$\overline{\eta_t\eta_{t+n\Delta t}}=\delta_{n,0}$, the  delta indicating the Kronecker symbol.  As discussed in Sect. \ref{sect:f} below, in typical cases the spurious drift \cite{risken} is negligible. We are interested in understanding if 
there exist two functions $D^{(1)}(Y)$ and $D^{(2)}(Y)$, and a time step $\Delta t$ for which the time evolution of $Y$ that one would simulate with an 
Euler algorithm displays the same moments $\overline{Y^k(t)}$, for any $k$, equal to 
$\overline{Y_t^k}$ at any time $t$, where $\overline{Y_t^k}$ are calculated from the trajectories 
generated by Eq. (\ref{eq:euler}). 
More specifically, one would like that the difference between the true moments and those 
calculated by Eq. (\ref{eq:euler}) goes to zero as $M\to\infty$.

In general, the time step $\Delta t$ of the effective equation (\ref{eq:euler}) can be different (i.e., larger) than the time step $\Delta t_{mic}$ of the underlying, full--dimensional simulation. The effective dynamics can thus be a coarse--graining in time of the original dynamics. A relevant question we shall ask is then what is the most suitable time step $\Delta t$ for which Eq. (\ref{eq:euler}) best reproduces the microscopic dynamics. It is not granted that the best choice is the smallest possible value, that is $\Delta t=\Delta t_{mic}$.

For example, one of the advantages of using a framework controlled by finite--difference dynamic equations, and in particular a large value of $\Delta t$, is that one  avoids the problems associated with the Markov--Einstein time scale \cite{Honisch:2011wv}. 
Moreover, since a typical problem associated with dimensional reduction is the non--Markovianity 
of the reduced dynamics, the use of a large time step $\Delta t$ increases the probability of 
ending up into a Markovian process.
If we are able to find a set of parameters for which Eq. (\ref{eq:euler}) reproduces the 
dynamics of the system projected on $Y$, then this is Markovian by definition.

Within this framework one can in principle obtain the drift and the diffusion coefficients at 
a given point $Y_0$. To do that, one generates $M$ trajectories starting from points $r$ of conformational space displaying $Y(r)=Y_0$ and evolving the system for a  time $\Delta t$. From Eq. (\ref{eq:euler}), moving $Y_t$ to the left--hand side of the equation, evaluating it in $t=0$ and averaging over the $M$ trajectories, one obtains
\begin{align}
 D^{(1)}(Y_0) &= \frac{ \overline{Y_{\Delta t}-Y_0} }{\Delta t} 
                 - \sqrt{ \frac{2D^{(2)}}{\Delta t} } \overline{\eta} \nonumber\\
 D^{(2)}(Y_0) &= \frac{ \overline{ [Y_{\Delta t}-Y_0]^2 } }{2\Delta t} 
                 + \frac{1}{2} \big[ D^{(1)}(Y_0) \big]^2\Delta t 
                 - D^{(1)} \big( \overline{Y_{\Delta t}-Y_0} \big) = \nonumber \\
              &= \frac{ \overline{ [Y_{\Delta t}-Y_0]^2 } }{2\Delta t} 
                 + \frac{1}{2} \big[D^{(1)}(Y_0) \big]^2 \Delta t
                 - D^{(1)}(Y_0)\bigg(D^{(1)}(Y_0) \Delta t 
                 + \sqrt{ 2D^{(2)}(Y_0)\Delta t} \cdot \overline{\eta}\bigg) = \nonumber \\
              &= \frac{ \overline{ [Y_{\Delta t}-Y_0]^2 } }{2\Delta t} 
                 - \frac{1}{2} \big[ D^{(1)}(Y_0) \big]^2 \Delta t
                 - D^{(1)}(Y_0)\sqrt{ 2D^{(2)}(Y_0) \Delta t} \cdot \overline{\eta}.
\label{eq:km2}
\end{align}
The last terms of both quantities go to zero, because $\overline{\eta}$ does, as $M\to\infty$. The straightforward way to obtain $D^{(1)}$ and $D^{(2)}$ is to generate $M$ short trajectories of duration $\Delta t$, from the first and last point of each trajectory one calculates the displacement $Y_{\Delta t}-Y_0$ and its square, and substitutes their averages in Eqs. (\ref{eq:km2}). Importantly, at variance with the expressions resulting from the standard Kramers--Moyal expansion, in Eqs. (\ref{eq:km2}) no short--time limit is necessary, but the the time increment $\Delta t$ is the same which defines the dynamic equation (\ref{eq:euler}).

\subsection{Calculation of the drift coefficient in a test model}

Unfortunately, the direct application of Eqs. (\ref{eq:km2}) is not straightforward, because of the error introduced by $\overline{\eta}$ for finite values of $M$. Consider for example the motion of a particle in a one--dimensional harmonic potential $k/2(Y-Y_c)^2$ characterized by  $D^{(1)}_{true}=-k(Y-Y_c)/\gamma$ and $D^{(2)}_{true}=k_BT/\gamma$, where  $\gamma$ is the friction coefficient and $k_BT$ is the thermal energy. 
This example was chosen because a particle in a one--dimensional harmonic potential is the simplest system displaying non--zero drift and diffusion coefficient, and is the first--order approximation of any attractive interaction that can be found in biological systems. Biomolecules typically display length scale of nanometers, energy scale comparable to the thermal energy at room temperature ($k_BT\sim 2.5$ kJ/mol), corresponding to forces of the order of tens of pN.
The numerical values of the simulations $k=-1$ pN/nm, $\gamma=10^{-11}$ Kg/s=$6\cdot 10^3$ kDa/ns, $k_BT=2.5$ kJ/mol, $Y_c=1$ nm (and thus $D^{(1)}_{true}=-0.1$ nm/ns; $D^{(2)}_{true}=0.4$ nm$^2$/ns) are chosen as typical orders of magnitude for biomolecules. 

\begin{figure*}
\includegraphics[scale=0.32]{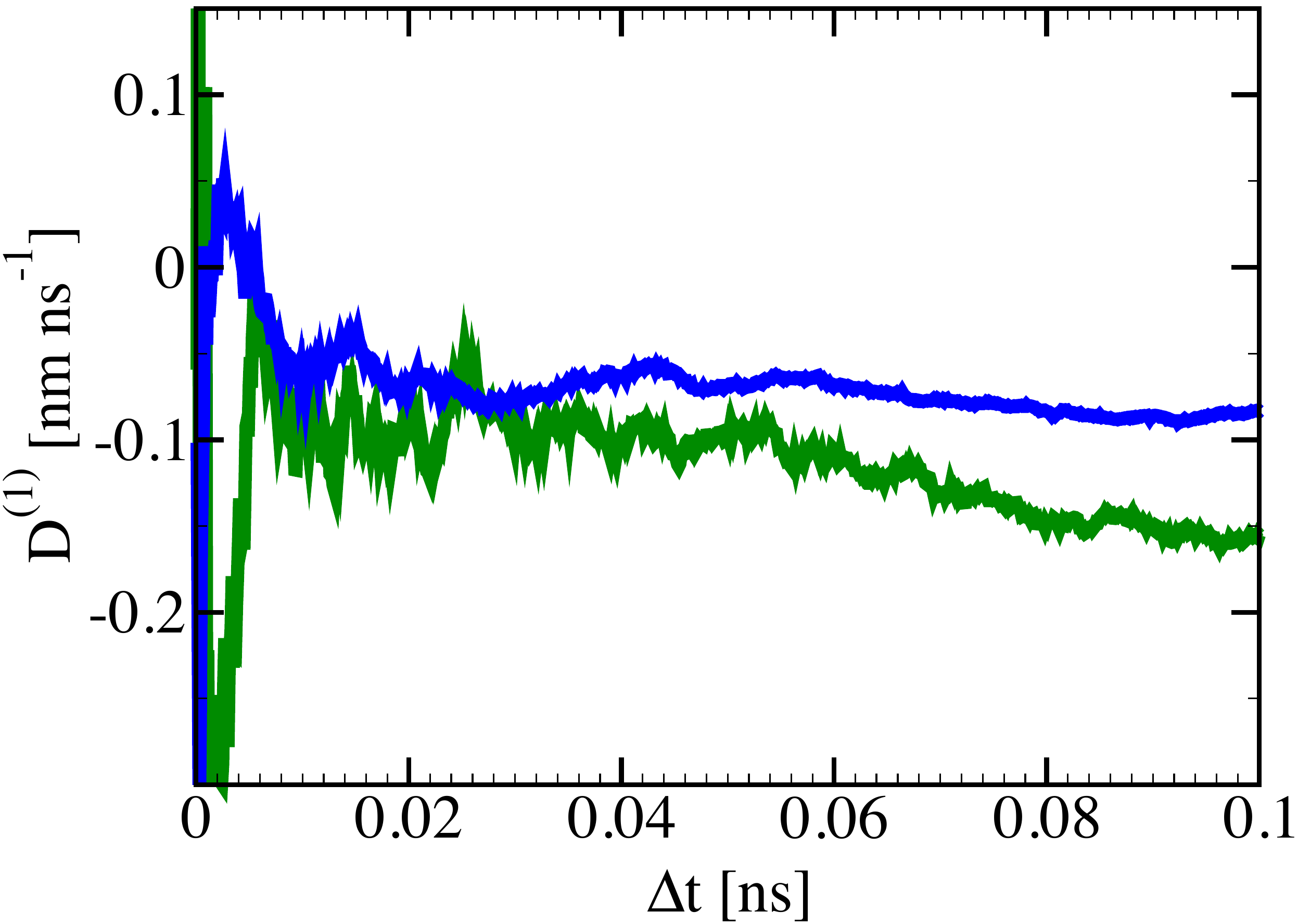}
\caption{The value of $D^{(1)}$ as a function of $\Delta t$ for the harmonic oscillator, 
calculated with Eq. (\protect\ref{eq:km2}) making use of different numbers $M$ of trajectories. 
The green (light gray) curve corresponds to $M=10^4$, the blue (dark gray) curve to $M=10^5$. The true value of $D^{(1)}$
is $-0.1$ nm/ns.}
\label{fig1}
\end{figure*}

In Fig. \ref{fig1} we display the value of $D^{(1)}$, obtained applying Eq. (\ref{eq:km2}) 
to different numbers of trajectories of length $0.1$ ns, generated by an overdamped 
Langevin equation and calculating for each step the value 
$(\overline{Y_{\Delta t}-Y_0})/\Delta t$.
It is apparent that for small values of $\Delta t$, comparable to $\Delta t_{mic}$, the 
curves are very noisy and do not allow for the determination of  $D^{(1)}$. 
The reason is that $\overline{\eta}$ is a stochastic number which goes to zero as $1/M^{1/2}$; the condition for noise to be small in the first line of  Eq. (\ref{eq:km2}) is then
\begin{equation}
M \gg  \frac{2D^{(2)}}{[D^{(1)}]^2\Delta t},
\label{eq:condM}
\end{equation}
which is stronger the smaller is $\Delta t$. This prevents the use of $\Delta t\approx\Delta t_{mic}$. In fact, the intrinsic time scale of a harmonic oscillator is $\gamma/k=Y_0/D^{(1)}$, and thus $\Delta t_{mic}$ must be chosen several orders of magnitude smaller than this time scale, say $h\cdot Y_0/D^{(1)}$ with $h\sim 10^{-4}$. The condition on the noise is then $M \gg 2D^{(2)} / h D^{(1)} Y_0$; since in typical systems the drift is comparable with noise $D^{(2)}\sim D^{(1)} Y_0$, one needs $M \gg 10^4$. 

Using a large $\Delta t$ reduces the random noise, because it corresponds effectively to averaging more displacements. In fact, one can think of a large time interval $\Delta t$ as composed of $m$ sub--intervals $\Delta t'$; the quantity $\overline{Y_{\Delta t}-Y_0}/\Delta t$ can thus be seen as an average over both the $M$ trajectories and the $m$ displacements calculated in each trajectory, for a total of $M\cdot m$ terms. The drawback is that only the  displacement in the first sub--interval of each trajectory starts exactly at $Y_0$, while the others are only approximations of this. The approximation worsen for larger $m$, and thus for larger $\Delta t$, and makes $D^{(1)}$ not a property of the very point $Y_0$, but an average of drift coefficients over a large number of points in the neighborhood of $Y_0$. As shown in  Fig.  \ref{fig1}, at large values of $\Delta t$ the calculated drift coefficient departs from its true value of $-0.1$ nm/ns. In other words, using a large $\Delta t$ one is trading a random error for a systematic error.
As a consequence, the direct use of Eq. (\protect\ref{eq:km2}) is impractical.

\subsection{The diffusion coefficient}

The situation is different for the diffusion coefficient. According to Eq. (\ref{eq:km2}), the expression for $D^{(2)}$ contains three terms. The one in the middle of the right--hand side depends on $D^{(1)}$ which, as discussed above, one is not able to calculate. This is however small if
\begin{equation}
[D^{(1)}]^2 \ll \frac{2 D^{(2)}}{\Delta t}.
\label{eq:cond}
\end{equation}
With the typical values listed above, 
using $\Delta t=10^{-4}$ ns this condition gives $|D^{(1)}|\ll 90$ nm/ns, corresponding to forces of the order of $10^3$ pN, 
which are huge on a biological scale. Thus it can be neglected in most cases.

The other source of error when using Eq. (\ref{eq:km2}) for $D^{(2)}$ is the noise which appears explicitly in the third line of Eqs. (\ref{eq:km2}). However it goes to zero as $\Delta t M^{-1/2}$, thus even at finite $\Delta t$ it is negligible for large $M$.

\begin{figure*}
\includegraphics[width=\textwidth]{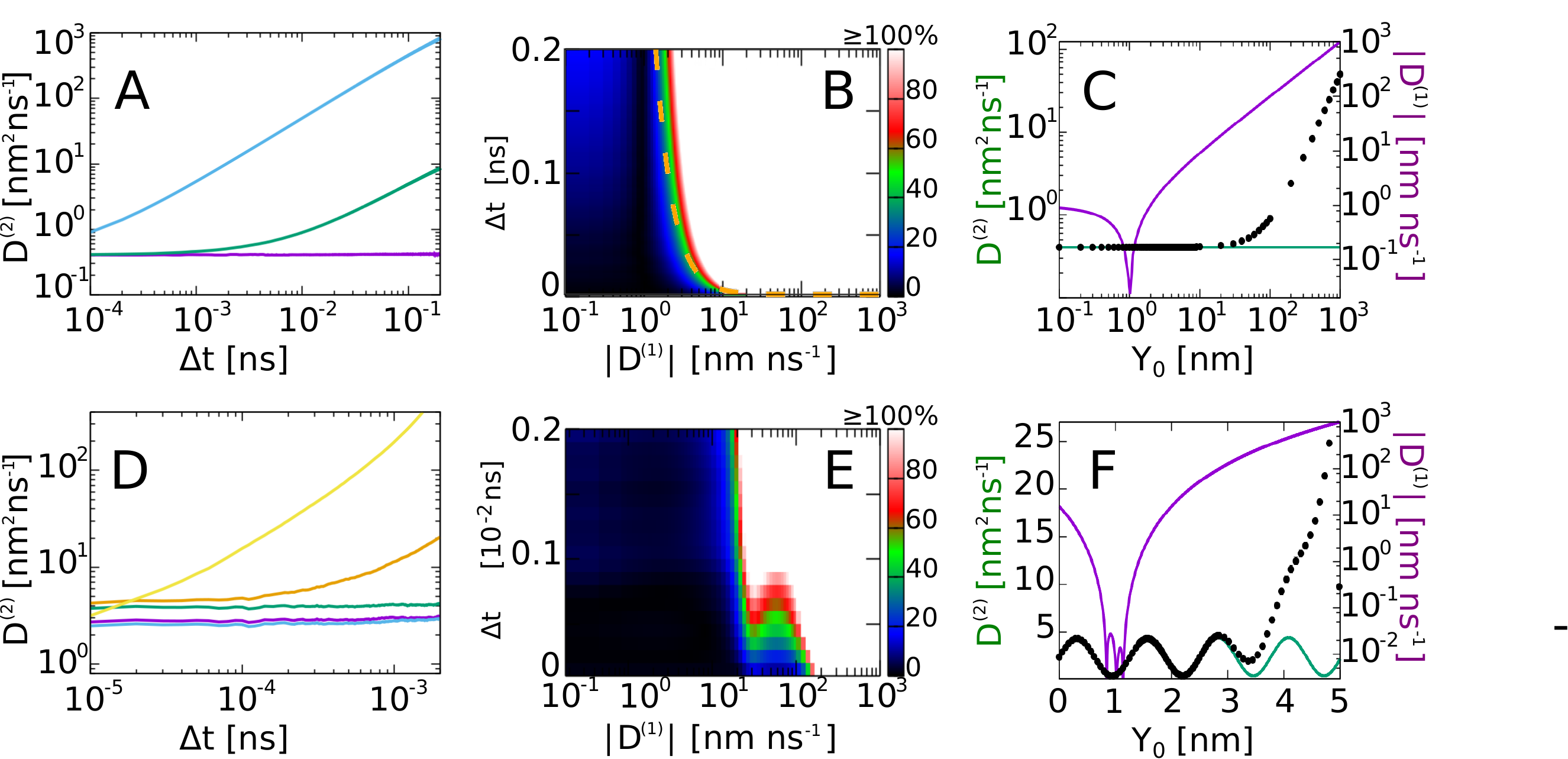}
\caption{For the harmonic spring, (A) the back--calculated $D^{(2)}$ as a function of $\Delta t$ in the cases $D^{(1)}_{true}=1$ nm/ns (purple (lower) line),  $D^{(1)}_{true}=10$ nm/ns (green (middle) line) and $D^{(1)}_{true}=10^2$ nm/ns (blue (upper) line). 
(B) The percentage error in the back--calculation with respect to the true value; the dashed curve indicates the threshold given by Eq. (\protect{\ref{eq:cond}}). 
(C) The values of $D^{(1)}_{true}$ (purple (upper) curve) and $D^{(2)}_{true}$ (green (lower) curve)  as a function of the elongation of the spring; the solid circles indicate the  back--calculated profile of $D^{(1)}$ using $\Delta t=\Delta t_{mic}$. 
(D), (E) and  (F) are the same as (A), (B) and (C), respectively, for the system displaying a two--state thermodynamics and a sinusoidal diffusion coefficient (cf. panel F). }
\label{fig:d2_definizione}
\end{figure*}

We have tested the calculation of $D^{(2)}$ from Eq. (\ref{eq:km2}), 
assuming $\overline{\eta}=0$, in two test cases. In the case of the harmonic spring 
already considered above, the quantity $\overline{(Y_{\Delta t}-Y_0)^2}/(2\Delta t)$ 
has been calculated from 0.2 ns simulations and 
the results are displayed in the upper panels of 
Fig. \ref{fig:d2_definizione}. 
As shown in Fig. \ref{fig:d2_definizione}A, if the drift is not too large 
($\lesssim 10$ nm/ns), $D^{(2)}$ as a function of $\Delta t$ is linear at small 
$\Delta t$, as expected from Eq. (\ref{eq:km2}), and assumes the correct value 
$0.400$ nm$^2$/ns (i.e., with an error smaller than 1\%). 

The accuracy  in the back--calculation of $D^{(2)}$ is displayed in Fig. \ref{fig:d2_definizione}B and the results are good if the drift is less than $\sim 10 $ nm/ns, while the prediction becomes unreliable for larger drifts. The reason for this loss of accuracy is in the approximation associated with neglecting the spurious term in Eq. (\ref{eq:km2}), that is the failure of condition (\ref{eq:cond}). The  threshold curve defined by Eq. (\ref{eq:cond}) is displayed with a dashed line in Fig.  \ref{fig:d2_definizione}B.
Thus, the maximum value of the drift that allows the calculation of the diffusion constant is for $\Delta t=\Delta t_{mic}$, and in this example assumes the value $D^{(1)}\approx 20$ nm/ns. This is anyhow a rather large drift coefficient, corresponding to a force of $\gamma D^{(1)}\sim 200$ pN. We can conclude that the use of $\Delta t=\Delta t_{mic}$ seems the best choice.

The above procedure is repeated starting from different initial points $Y_0$ using $\Delta t=\Delta t_{mic}$ ($=10^{-4}$ ns), and the back--calculated values of $D^{(2)}$ are displayed with solid circles in Fig.  \ref{fig:d2_definizione}C, superposed to the curve which defines $D^{(2)}_{true}$. Again, the results are good up to the value of $Y_0$ at which $D^{(1)}_{true}$ (purple curve) is of the order of $10$ nm/ns.
 
Another test was carried out with a more challenging one--dimensional system, which  is bistable (cf. purple curve in Fig.  \ref{fig:d2_definizione}F) and in which the diffusion coefficient is sinusoidal (cf. green curve in  Fig.  \ref{fig:d2_definizione}F).  This  shapes do not have direct relevance to describe any specific biological system, but it was chosen just because of its high complexity, following the idea that if the algorithm works for this case, it will work also for simpler, biologically--inspired force fields and diffusion coefficients.  A time step $\Delta t_{mic}=10^{-6}$ ns was used to generate the trajectories of  length $2\cdot 10^{-3}$ ns.  The results are similar to the case of the harmonic spring, that is one can back--calculate  the diffusion coefficient with good accuracy (see Figs. \ref{fig:d2_definizione}D, E and F) provided that the true drift coefficient is small enough  that the system can diffuse within the time $\Delta t$.

\subsection{Higher Kramers--Moyal coefficients}

From the Langevin--like equation (\ref{eq:euler}) one can also find the higher moments, like
\begin{equation}
\overline{(Y_{\Delta t}-Y_0)^3}-( D^{(1)} \Delta t)^3 -6D^{(1)}D^{(2)}\Delta t^2 = z\cdot\overline{\eta} +  w\cdot \overline{\eta^3} ,
\label{eq:m3}
\end{equation}
where $z=18^{1/2}(D^{(1)})^2(D^{(2)})^{1/2}\Delta t^{5/2}$ and $w=(2D^{(2)}\Delta t)^{3/2}$.
In the limit of small $\Delta t$, the left--hand side of this equation corresponds to the third Kramers--Moyal coefficient multiplied by $\Delta t$, so by extension we will label it as $D^{(3)}\Delta t$. In fact, combining Eq. (\ref{eq:m3}) with the properties of $\overline{\eta}$ it follows that $D^{(3)}$ should vanish as $z/M^{1/2}$ for Eq (\ref{eq:euler}) to hold. If this is the case, all higher moments should vanish by Pawula theorem \cite{risken}. 

\begin{figure*}
\includegraphics[scale=0.5]{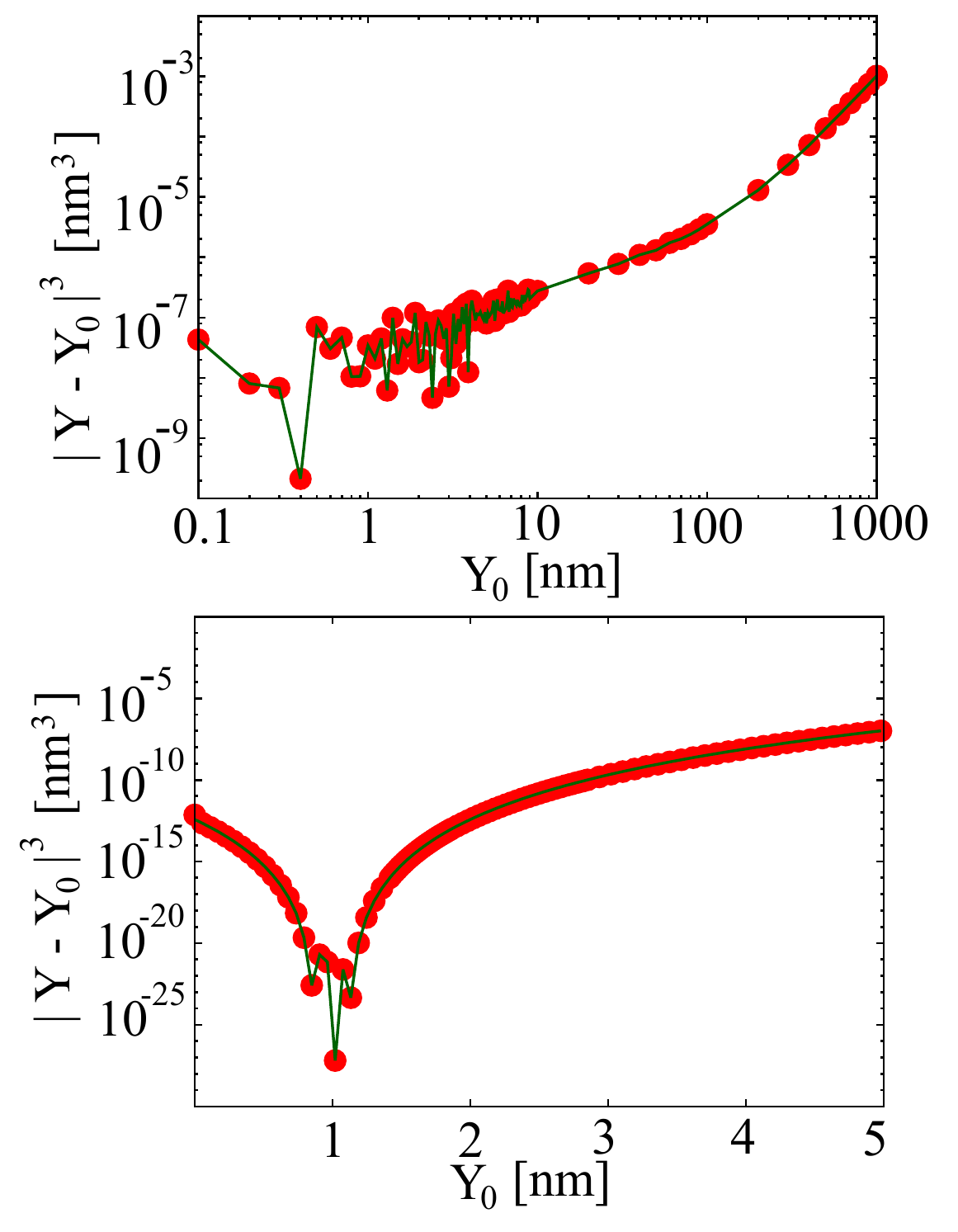}
\caption{The value of the third moment of the displacement, calculated for the harmonic spring (above) and for the two-state system with sinusoidal diffusion coefficient (below).}
\label{newfig3}
\end{figure*}

The validity of  Eq. (\ref{eq:m3}) is difficult to establish numerically, because of the problems already discussed in estimating $D^{(1)}$. Consequently, we only checked the  condition
\begin{equation}
\overline{(Y_{\Delta t}-Y_0)^3}  \ll [  D^{(2)} \Delta t]^{3/2} \;\;(=w)
\label{eq:d3}
\end{equation}
which means that the skewness of the displacements is negligible with respect to the diffusion in each run for small $\Delta t$. With the typical values of the diffusion coefficients we chose, $w\sim 10^{-7}$ nm$^3$. This condition is satisfied both in the case of the particle in the harmonic well and of the bistable system, except for extreme choices of the drift coefficient (cf. Fig. \ref{newfig3}).

\section{Another strategy to calculate Drift and Diffusion coefficients}

\subsection{Drift and diffusion coefficients from an iterative approach}

An alternative way to calculate the drift and diffusion coefficients in a point of the conformational space of the system is by a locally linear approximation of the force and a constant approximation of the diffusion coefficient in the neighborhood 
of that point. Writing the force locally as $f(Y)=k(Y-Y_c)$, the drift coefficient is
\begin{equation}
D^{(1)}(Y)=\rho (Y-Y_c),
\label{eq:d11}
\end{equation}
where $\rho=k/\gamma$. Defining $B=1+\rho\Delta t$, the Euler version of Langevin equation can thus be iterated to give
\begin{align}
Y_{\Delta t}  &= BY_0 - (B-1)Y_c + (2D^{(2)}\Delta t)^{1/2}\eta_0;\\
Y_{2\Delta t} &= BY_{\Delta t} - (B-1)Y_c + (2D^{(2)}\Delta t)^{1/2}\eta_{\Delta t}=\nonumber\\
              &= B^2Y_0 - (B-1)(B+1)Y_c + 
                 (2D^{(2)}\Delta t)^{1/2}(B\eta_0+\eta_{\Delta t});\nonumber\\
Y_{3\Delta t} &= B^3Y_0 - (B-1)(B^2+B+1)Y_c +\\
              &\hspace{15pt}+ (2D^{(2)}\Delta t)^{1/2}(B^2\eta_0+B\eta_{\Delta t}+\eta_{2\Delta t});\nonumber\\ 
...
\end{align}
which can be generalized to $n$ steps by performing the geometric sum as
\begin{align}
Y_{n\Delta t}-Y_0&=(B^n-1)(Y_0-Y_c)+ \nonumber\\
&+ (2D^{(2)}\Delta t)^{1/2}\sum_i^{n-1}B^i\eta_{(n-i-1)\Delta t}.
\label{eq:iteratus}
\end{align}

The drift coefficient at point $Y_0$, or better the two parameters $\rho$ and $Y_c$ which define it, can be found from the average displacement of $Y$, that is
\begin{align}
\frac{\overline{Y_{n\Delta t}-Y_0}}{n\Delta t}&=\frac{(1+\rho\Delta t)^n-1}{n\Delta t}(Y_0-Y_c)+\nonumber\\
&+\left(\frac{2D^{(2)}}{\Delta t}\right)^{1/2}\frac{1-(\rho\Delta t)^n}{1-\rho\Delta t}\frac{1}{n}\sum_{i=0}^{n-1}\overline{\eta}_i.
\label{eq:drift}
\end{align}
Now we expect the term containing $\overline{\eta}$ to be small, because its standard deviation goes to zero as $(M\times n)^{-1/2}$, and because even if $\Delta t$ is small, the prefactor $[D^{(2)}/\Delta t]^{1/2}$ is much smaller than in Eq. (\ref{eq:km2}). For example, if one considers $D^{(2)}\sim D^{(1)}Y_0$ and $\Delta t\sim 10^{-4} Y_0/D^{(1)}$, the noise term results only of the order of $10^2\cdot(M\times n)^{-1/2}$.

Incidentally, neglecting  $\overline{\eta}$, Eq. (\ref{eq:drift}) converges in the limit $n\Delta t\to 0$ to $\rho(Y_0-Y_c)$ which one would expect in the standard Langevin differential equation.

Thus, neglecting $\overline{\eta}$ in Eq. (\ref{eq:drift}) and defining $\tau=n\Delta t$ one can write
\begin{equation}
K(\tau)\equiv\overline{Y_{\tau}-Y_0}=[(1+\rho\Delta t)^{\tau/\Delta t}-1](Y_0-Y_c).
\label{eq:dy2fit}
\end{equation}
The left--hand side of this expression is a function of elapsed time $\tau$ which can be obtained from the simulations, while the right--hand side depends parametrically on $\rho$, $Y_c$  and $\Delta t$.

Similarly to the calculation of the drift coefficients in Eq. (\ref{eq:drift}) it is possible to use the iterative procedure described above to obtain the diffusion coefficient. In fact from Eq. (\ref{eq:iteratus}) one can obtain
\begin{align}
 \overline{\left( Y_{\tau}-Y_0   \right)^2 }&= \left[  ( 1+\rho\Delta t)^{\tau/\Delta t} -1  \right]^2 (Y_0-Y_c)^2 - 2 D^{(2)}
\frac{ 1- (1+\rho\Delta t)^{2\tau/\Delta t} }{\rho(2+\rho\Delta t)},
\end{align}
which can be combined with Eq. (\ref{eq:dy2fit}) to give
\begin{align}
J(\tau)\equiv \overline{\left( Y_{\tau}-Y_0   \right)^2 } - \left( \overline{Y_{\tau}-Y_0} \right)^2  &=    2D^{(2)}
\frac{  (1+\rho\Delta t)^{2\tau/\Delta t} -1}{\rho(2+\rho\Delta t)},
\label{eq:dy3fit}
\end{align}
which depends parametrically on $D^{(2)}$, $\Delta t$ and $\rho$.

\subsection{Determination of the parameters}

The idea is thus to fit the numerical values of $K(\tau)$ and $J(\tau)$, obtained 
from the MD simulations previously generated, with the expressions of Eqs. (\ref{eq:dy2fit}) 
and (\ref{eq:dy3fit}) to obtain the four parameters $\rho$, $Y_c$ 
(which define  $D^{(1)}$),  $D^{(2)}$ and, in principle, $\Delta t$. 

\begin{figure*}
\includegraphics[scale=0.32]{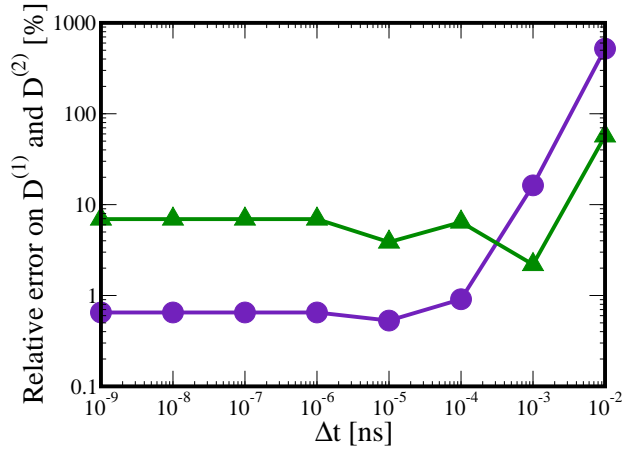}
\caption{The quality of the reconstruction of $D^{(1)}$ (purple circles) and $D^{(2)}$ (green triangles) poorly depends on $\Delta t$, provided that the two functions $J(\tau)$ and $K(\tau)$ have sufficient temporal resolution to be fitted.}
\label{newfig4}
\end{figure*}

The quantity $\Delta t$ sets the time scale of the time--dependent parameters, 
in the sense that the equations are invariant under the transformations 
$\rho\to\rho\Delta t$, $\tau\to\tau/\Delta t$ and $D^{(2)}\to D^{(2)}\Delta t$. 
The choice of $\Delta t$ is then not critical, provided that it is small 
enough to allow a high--resolution determination of the curves $J(\tau)$ and $K(\tau)$ 
(cf. Fig. \ref{newfig4}). 
Empirically, we find that the choice $\Delta t=10\Delta t_{mic}$ 
provides a good balance between the resolution of $K(\tau)$ and $J(\tau)$ 
and the feasibility of the fit. 

The shape of  $J(\tau)$ is usually curved (cf. solid curves in 
Figs. \ref{fig:model_fit_easy}A and \ref{fig:model_fit_hard}A), 
and so we expect it to be specified by (at least) three parameters; 
since the theoretical expression (\ref{eq:dy3fit}) for $J(\tau)$ depends on three parameters, we then expect to be able to fit them, and in particular to obtain $D^{(2)}$ and $\rho$ from it. 
On the other hand, Eq. (\ref{eq:dy2fit}) can be written as 
$\log K=\log(Y_0-Y_c)+\log (...)$, we can then fit $Y_c$ from 
the small--$\tau$ region of $\log K(\tau)$, which is in general rather 
flat (cf. dashed curves in Figs. \ref{fig:model_fit_easy}A and \ref{fig:model_fit_hard}A).

\subsection{Results for the test models}

\begin{figure*}
\includegraphics[width=\textwidth]{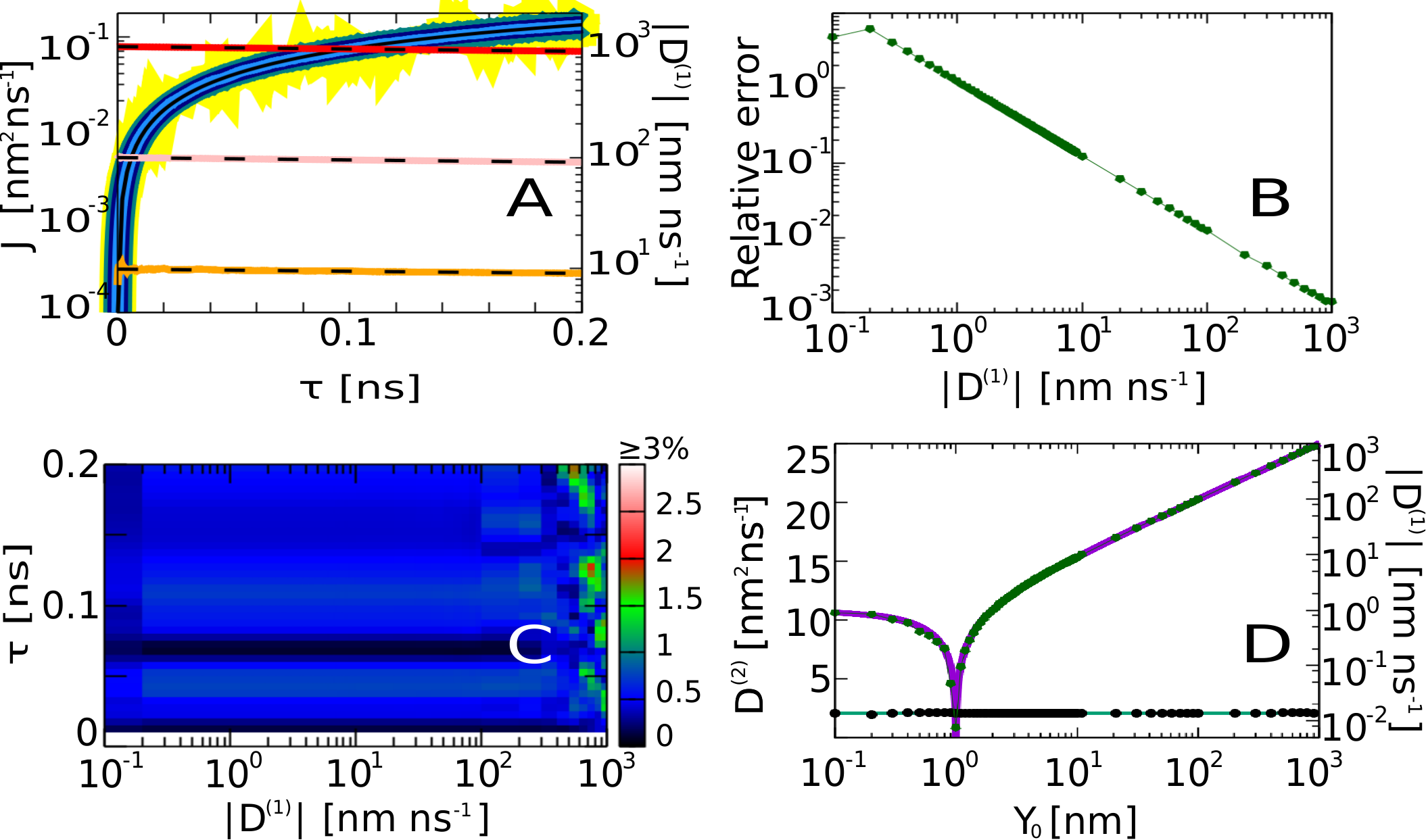}
\caption{
(A) The values $J$ (solid black curves) and $|D^{(1)}|$ (dashed black curves) as a  function of $\tau$ for the spring chain. The colored curves are the fits by  Eqs. (\protect\ref{eq:dy2fit}) and (\protect\ref{eq:dy3fit}), respectively. 
(B) The percentage standard error in the determination of $D^{(1)}$. 
(C) The percentage standard error in the determination of $D^{(2)}$ as a function of $\tau$  and $D^{(1)}_{true}$. 
(D) the reconstruction (green (upper) and black (lower) dots, respectively) of the true behavior of $D^{(1)}$ and $D^{(2)}$ (solid curves).}
\label{fig:model_fit_easy}
\end{figure*}

\begin{figure*}
\includegraphics[width=\textwidth]{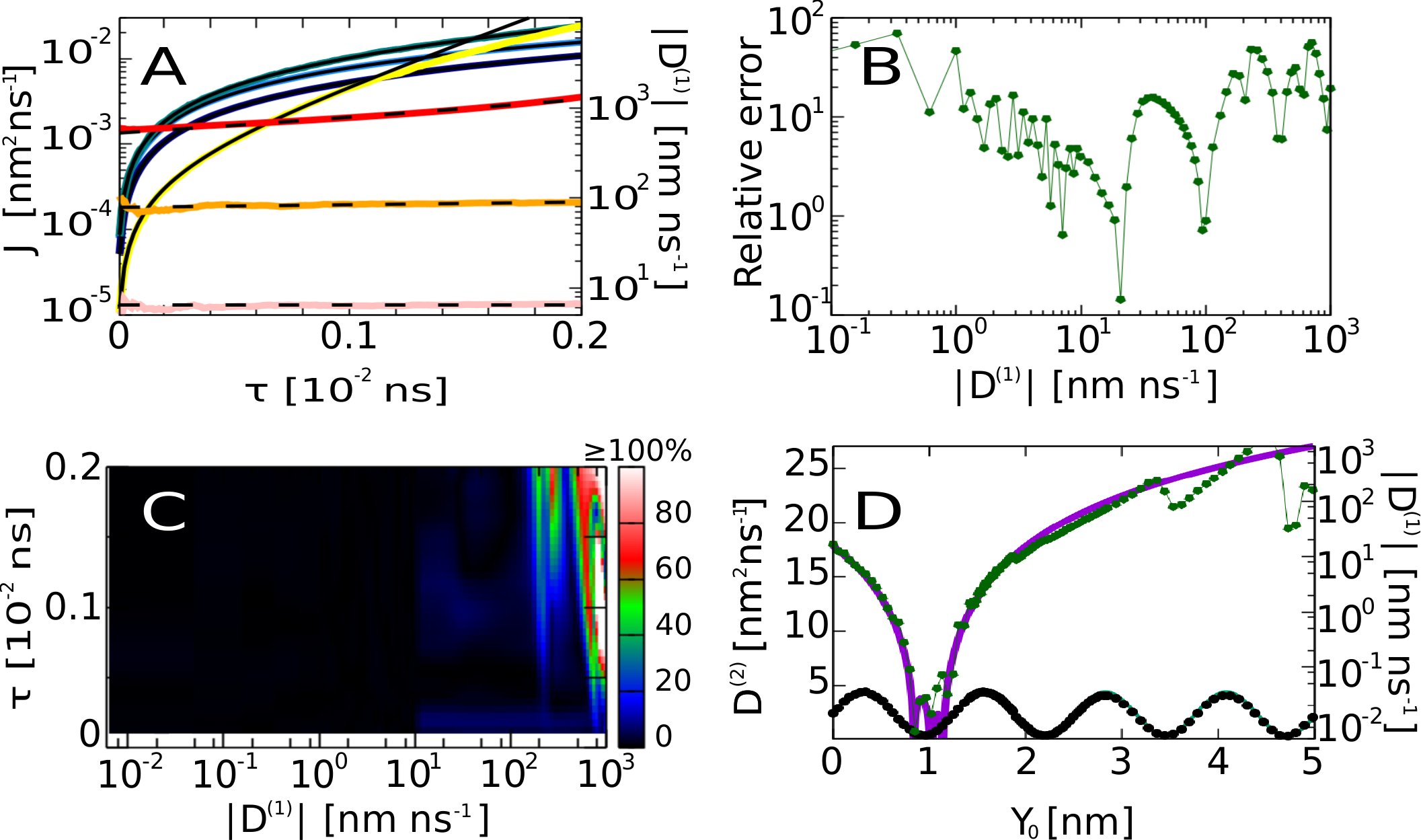}
\caption{The same as Fig. \protect\ref{fig:model_fit_easy} for a system displaying a two--state thermodynamics and a sinusoidal diffusion coefficient (cf. lower panels of Fig. \protect\ref{fig:d2_definizione}). }
\label{fig:model_fit_hard}
\end{figure*}

In Fig. \ref{fig:model_fit_easy} we report the back--calculation of the drift 
coefficient $D^{(1)}$ for the test case of a one--dimensional particle in a harmonic 
potential which, in spite of its simplicity, could not be achieved directly by definition (cf. Fig. \ref{fig1}). 
In Fig. \ref{fig:model_fit_easy}B it is shown the percentage standard error 
in the determination of $D^{(1)}$; the error is lower than 10\% in all cases 
and decreases for larger values of $D^{(1)}$, where the drift dominates over diffusion. 
Fig. \ref{fig:model_fit_easy}C shows the percentage standard error in the 
back--calculation of $D^{(2)}$. The error is less than 1\%. Good results 
were obtained calculating $D^{(2)}$ by its definition (cf. Fig. \ref{fig:d2_definizione}) 
at small values of $D^{(1)}_{true}$; the present method extends those results 
to any biologically--relevant value value of $D^{(1)}_{true}$. The overall 
reconstruction of the profile of $D^{(1)}$ and $D^{(2)}$ is displayed in 
Fig. \ref{fig:model_fit_easy}D.

Similar results were obtained for the more challenging system displaying 
a two--state thermodynamics and a sinusoidal diffusion coefficient 
(cf. lower panels of Fig. \ref{fig:d2_definizione}). In Fig. \ref{fig:model_fit_hard} 
we report the fits (panel A), the percentage standard error on $D^{(1)}$ 
(panel B) and $D^{(2)}$ (panel C), and the  reconstruction of the profile of 
the two coefficients (panel D). Overall, $D^{(1)}$ can now be calculated with 
good accuracy, and $D^{(2)}$ with a better accuracy than using Eq. (\ref{eq:km2}).

\section{Evaluation of the properties of collective variables} \label{sect:prot}

The final goal of the present work is to investigate whether some of the 
CVs commonly used to describe the dynamics of protein models 
are good reaction coordinates, that is whether we can describe the dynamics of 
the system in reduced dimensions by Eq. (\ref{eq:euler}). 
This is possible if the CV identifies uniquely $D^{(1)}$ and $D^{(2)}$, 
and if higher Kramers--Moyal coefficients are zero. 

The strategy we pursuit to challenge a CV $Y$ is that of 
generating a set of microscopic coordinates with the same value of $Y$, 
and calculating the distribution of $D^{(1)}$, $D^{(2)}$ and $D^{(3)}$ 
associated with that set. If $Y$ is a good reaction coordinate, such distributions 
should be strongly peaked, identifying a single value for $D^{(1)}$, $D^{(2)}$, 
and should be zero for $D^{(3)}$. The standard deviation $\sigma_1$ and $\sigma_2$ 
of the distributions $D^{(1)}$ and $D^{(2)}$, respectively, and the root mean 
square difference $\sigma_3$ from zero of $D^{(3)}$ can be regarded as measures 
of the quality of the CV as reaction coordinate. 
An ideal reaction coordinate should display $\sigma_1=\sigma_2=\sigma_3=0$.

We expect the quality of CVs to be structure dependent. 
As a consequence, we analyze separately the basic structural units of proteins, 
that is $\alpha$--helices and $\beta$--hairpins, in particular the second 
hairpin of GB1 domain and its helix \cite{Kraulis:1991ug} (pdb code: 1PGB).
For the same reason, we use a structure--based, implicit--solvent 
potential \cite{Noel:2010jf} which makes calculations particularly fast. 
The potential is the sum of atom--atom terms displaying a Lennard-Jones shape with 
a minimum at energy $-1$ (in arbitrary energy units) at the native distance of the pairs of atoms.
All the trajectories are generated using Gromacs 5.1\cite{Abraham:2015}, at 
a low temperature ($T=0.92$ in energy units), at which the native conformations are stable; 
the post-processing of data is done with Plumed 2.3\cite{Tribello:2014}.

We focused our attention on two popular CVs used in protein folding, namely the distance RMSD 
(dRMSD) to the native conformation $\{r_i^N\}$,
\begin{equation}
dRMSD(\{r_i\})=\left[ \frac{1}{N(N-1)}\sum_{ij}\left( |r_i-r_j| - |r^N_i-r^N_j|\right)^2 \right]^{1/2}
\end{equation}
and the fraction of native contacts $q$,
\begin{equation}
 q = \frac{1}{W} \sum_{ij \in C }^{W} s(r_{ij}) \qquad\text{with }\qquad s(r)=
\begin{cases}
    1                                               & \text{if } x\leq d_0\\
    1 - \tanh \left( \frac{ r - d_0 }{ r_0 }\right) & \text{otherwise} 
\end{cases}
\end{equation}
where the sum is performed on the $W$ pairs  of atoms $i$ and $j$ belonging to the set $C$ of pairs displaying 
in the native conformation distances $r_{ij}<5$ \AA\ and $|i-j|\ge 4$. We have used $r_0=1$ \AA\ and $d_0=5$ \AA\ in the above definition.
For the two systems under study, we got $W=149$ for the $\alpha$--helix, $244$ for the $\beta$--sheet.

To calculate $D^{(1)}$ and $D^{(2)}$ we first generated a long trajectory (100ns) at 
$T=0.92$ from which we extracted $4\cdot 10^4$ frames and for each of them we calculated 
the dRMSD and $q$. From them, we extracted five subsets of 500 conformations each, 
displaying dRMSD in the ranges $[0.20,0.21]$nm, $[0.25,0.26]$nm, $[0.30,0.31]$nm, 
$[0.35,0.36]$nm, $[0.40,0.41]$nm, 
and other five subsets with $q$ in the ranges $[0.10-0.13]$, $[0.30-0.33]$, $[0.50-0.53]$, $[0.70-0.73]$, $[0.90-0.93]$. 
Every frame was the starting conformation for $M=300$ independent simulations, in which the corresponding CV was monitored.  From them, the functions $K$ and 
$J$ could be drawn and fitted and the values of $D^{(1)}$ and $D^{(2)}$ were calculated according to Eq.  (\ref{eq:d11}), (\ref{eq:dy2fit}) and (\ref{eq:dy3fit}).

\begin{figure*}
\includegraphics[width=\textwidth]{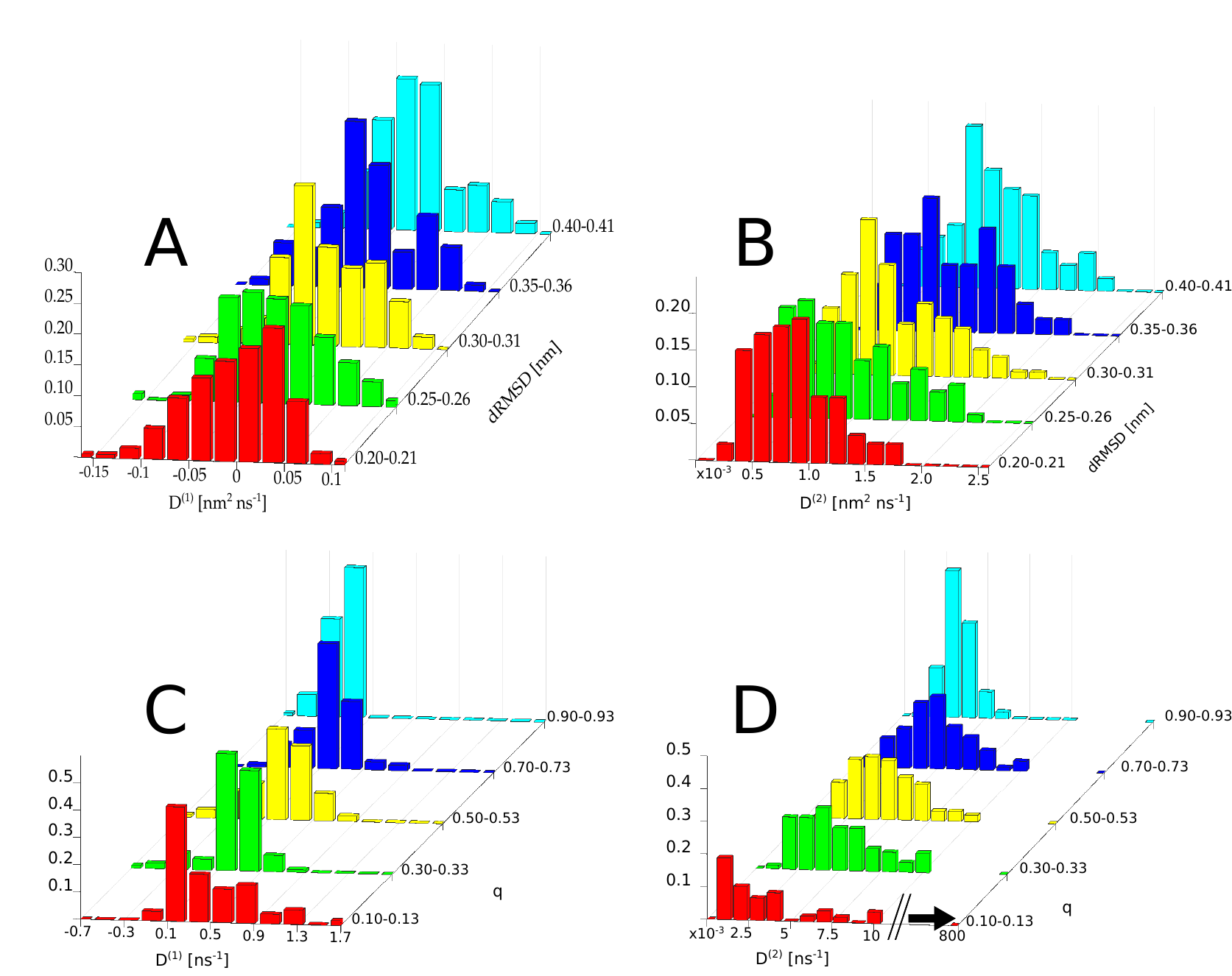}
\caption{ 
The distributions of drift (A--C) and diffusion (B--D) coefficients for the CVs $dRMSD$ 
(upper panels) and $q$ (lower panels), calculated in different intervals of the CV for 
the case of the $\alpha$--helix.}
\label{fig:helix_histo}
\end{figure*}

\begin{figure*}
\includegraphics[width=\textwidth]{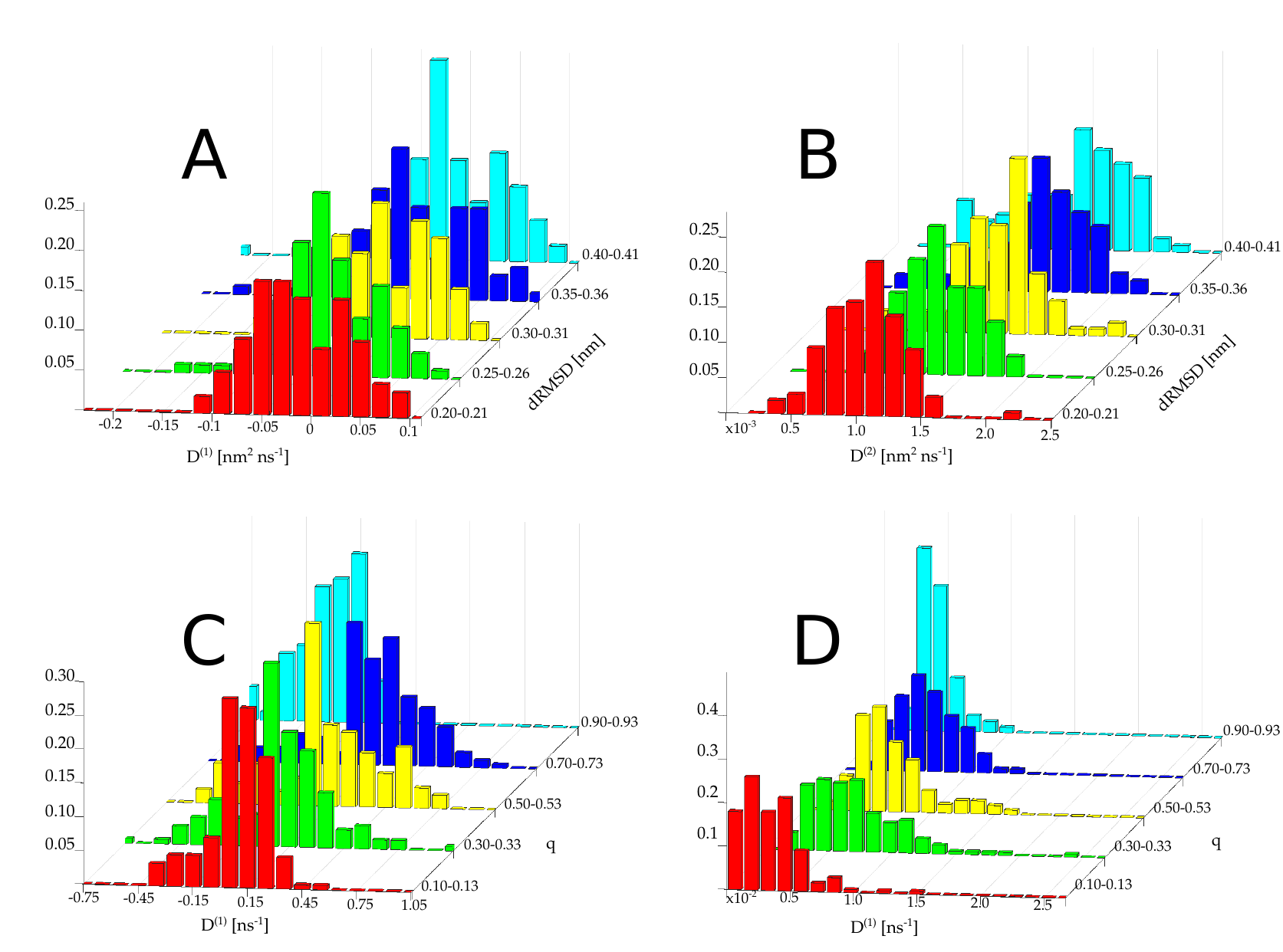}
\caption{The distributions of drift (A--C) and diffusion (B--D) coefficients for the CVs $dRMSD$ 
(upper panels) and $q$ (lower panels), calculated in different intervals of the CV for the case of the $\beta$--hairpin.}
\label{fig:hairpin_histo}
\end{figure*}

The distributions of coefficients $D^{(1)}$ and $D^{(2)}$ calculated for the 
$\alpha$--helix and the $\beta$--hairpin are displayed in Figs. \ref{fig:helix_histo} 
and \ref{fig:hairpin_histo}, respectively. The associated means and standard deviations 
are displayed in Figs. \ref{fig:helix_plots} and \ref{fig:hairpin_plots}, respectively. 
The value of $D^{(3)}$ is always negligible according to the criterion of Eq. (\ref{eq:d3}), the ratio between  $D^{(3)}\Delta t$ and $|D^{(2)}\Delta t|^{3/2}$ being of the order of $10^{-3}$ for the case of dRMSD and $10^{-2}$ for the case of $q$. 

In the case of the dRMSD,  both $\alpha$--helix and $\beta$--hairpin display a 
distribution of $D^{(1)}$ which is highly spread out, with standard deviations 
which are several times the mean. The mean is negative for all values of the dRMSD, 
but is decreasing for the $\alpha$--helix, while it is overall flatter for the 
$\beta$--hairpin. The diffusion coefficient is better defined by the dRMSD, the 
standard deviations being of the order of half of the mean. It is lowest at low 
dRMSD, where the conformational space is narrower, and is always increasing in the 
case of $\alpha$--helix, while it slightly decreases at large dRMSD in the case 
of $\beta$--hairpin. The standard deviation of both  $D^{(1)}$ and  $D^{(2)}$ are 
rather independent on the dRMSD.

The fraction of native contacts $q$ displays a drift coefficient whose standard 
deviation is of the order of the mean. It is positive except at $q\approx 0.9$, 
indicating that the equilibrium state has $0.7<q<0.9$. As in the case of dRMSD, 
the diffusion coefficient of the $\alpha$--helix decreases monotonically to the 
native state, while it has a bell--shaped behavior for the $\beta$--hairpin. 
The standard deviation of  $D^{(2)}$ relative to its mean is overall comparable 
with that of the dRMSD but, at variance with this, displays a weak decrease when 
moving to the native state.

\section{Effective force and dynamics in reduced dimension} \label{sect:f}

If a CV $Y$ is a reaction coordinate, then the effective force exerted on the system is opposite of the gradient of the free energy, calculated as a function of $Y$ \cite{risken}. In the formalism of Eq. (\ref{eq:red-langevin}), the effective force is $f=\gamma D^{(1)}$, where $\gamma$ is the effective friction coefficient given by Einstein's relation  $\gamma=T/ D^{(2)}$. The effective force associated with spurious drift \cite{risken} is negligible, being two orders of magnitude smaller than the drift coefficient.  A relevant question is then to which extent the effective force is equal to the opposite derivative of the free energy for the two variables studied in Sect. \ref{sect:prot}.

\begin{figure*}
\includegraphics[width=\textwidth]{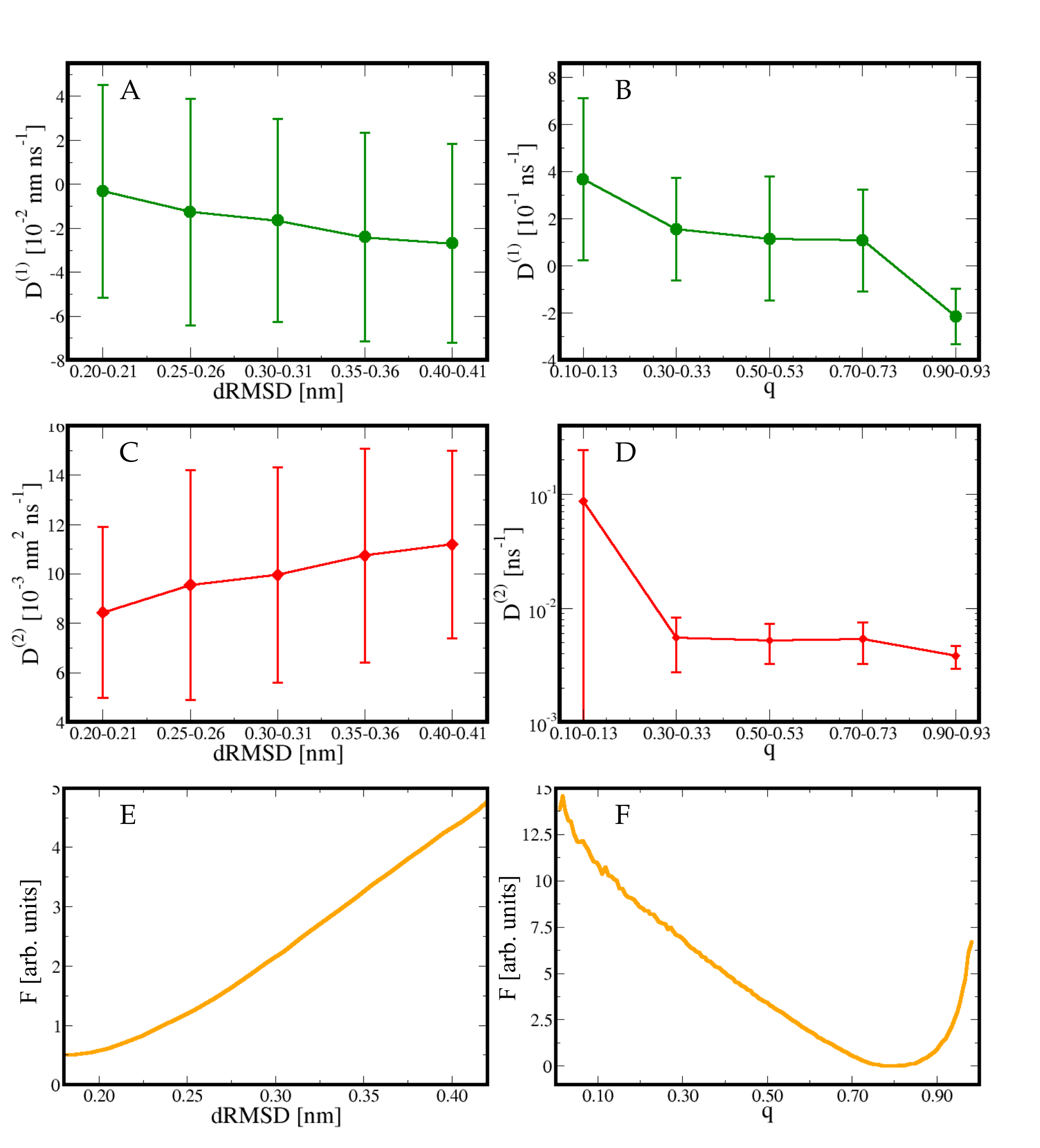}
\caption{The average and the standard deviation, plotted as error bar, 
of the drift (A--B) and diffusion coefficient (C--D) for the $\alpha$--helix, calculated 
in different intervals of dRMSD (left panels) and $q$ (right panels). Below, the free energies in the same region 
of interest for both dRMSD (E) and $q$ (F), at the simulation temperature $T=0.92$ (in energy units).}
\label{fig:helix_plots}
\end{figure*}

\begin{figure*}
\includegraphics[width=\textwidth]{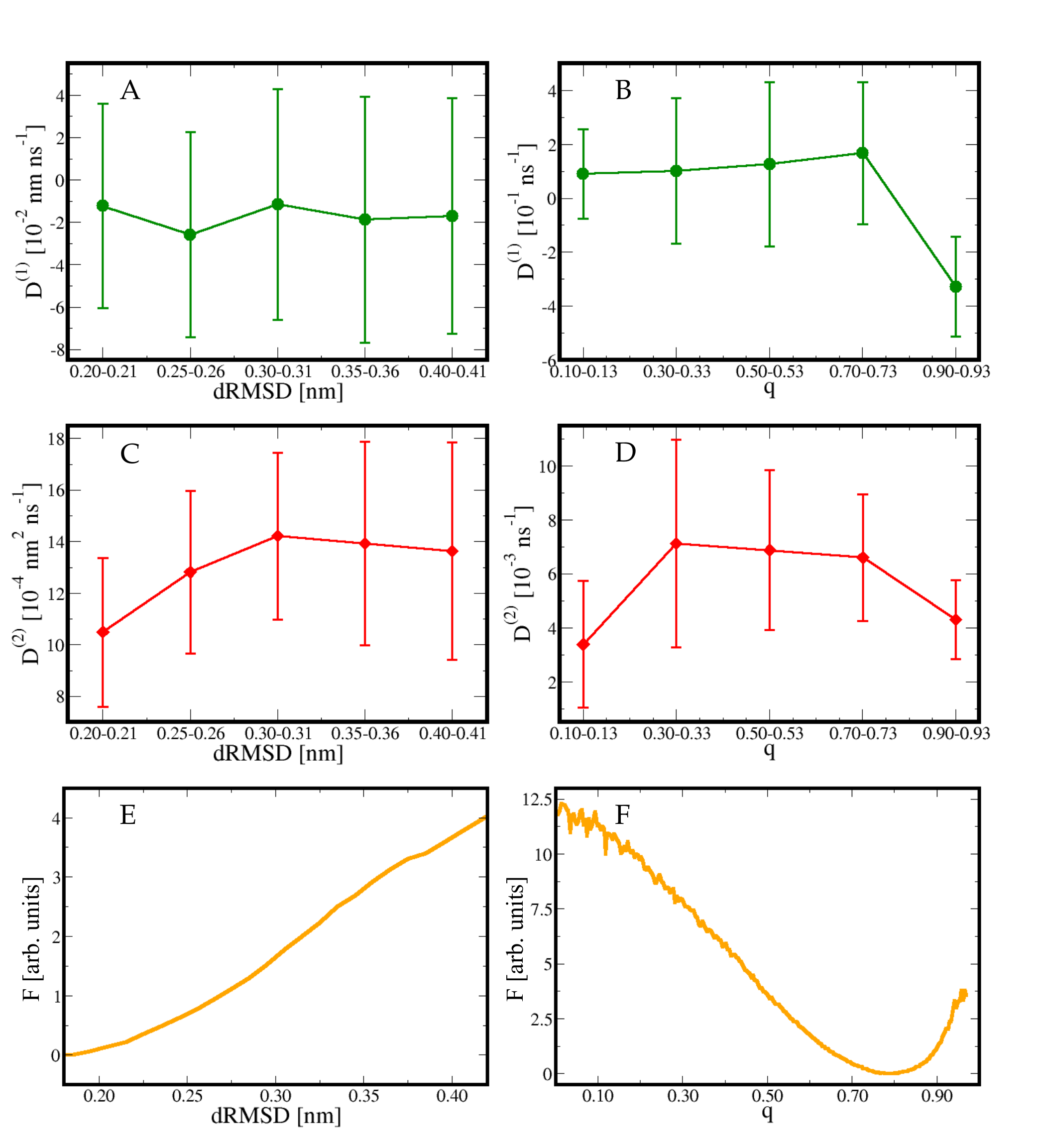}
\caption{
The average and the standard deviation, plotted as error bar, 
of the drift (A--B) and diffusion coefficient (C--D) for the $\beta$--hairpin, calculated 
in different intervals of dRMSD (left panels) and $q$ (right panels). Below, the free energies in the same region 
of interest for both dRMSD (E) and $q$ (F), at the simulation temperature $T=0.92$ (in energy units).}
\label{fig:hairpin_plots}
\end{figure*}

\begin{figure*}
\includegraphics[width=\textwidth]{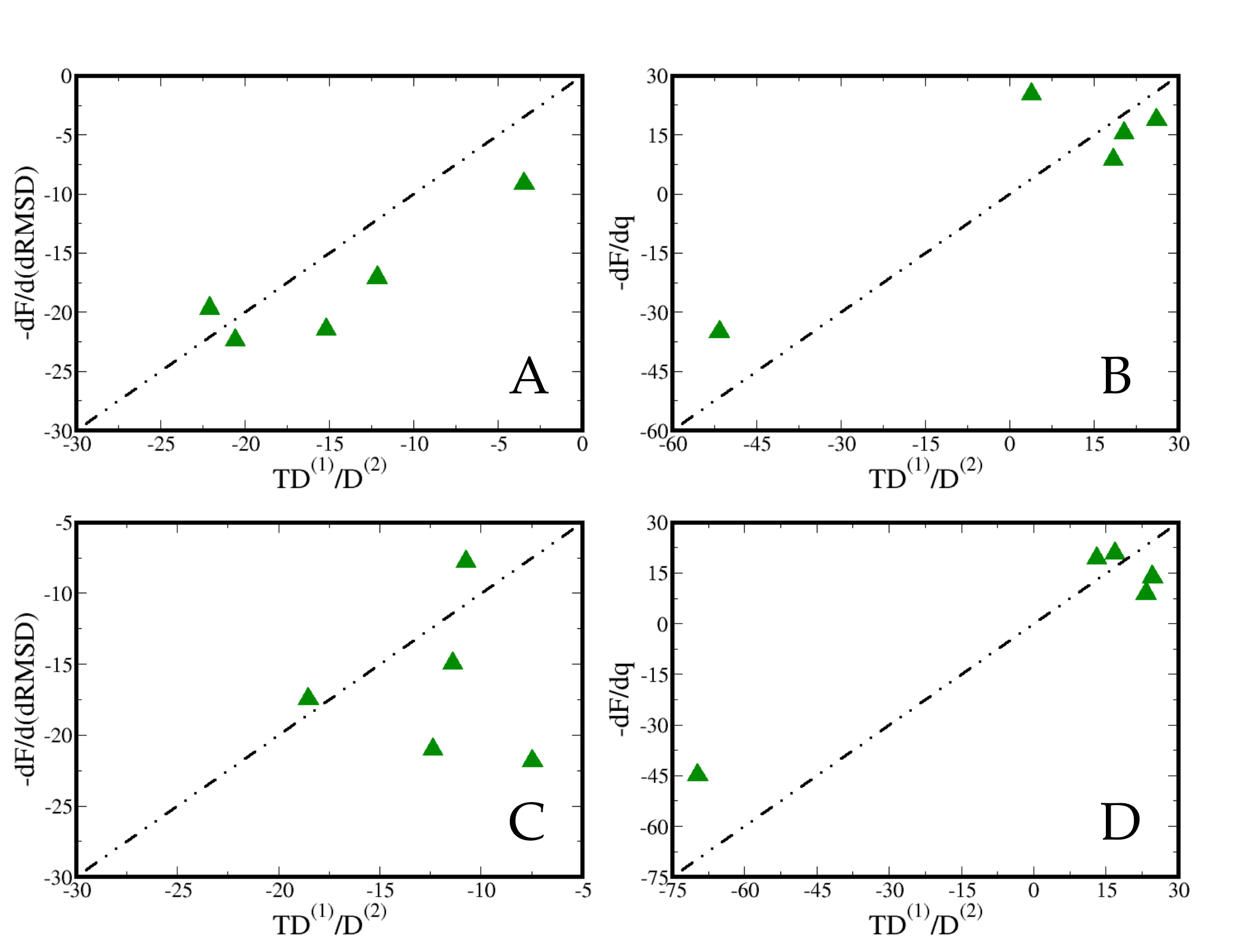}
\caption{
A comparison between effective force $TD^{(1)}/D^{(2)}$ obtained from the drift  and diffusion coefficients  and that obtained differentiating the equilibrium free energy  (cf. Figs. \protect\ref{fig:helix_plots} and \protect\ref{fig:hairpin_plots}) in the case of the $\alpha$--helix using the dRMSD (A)  and  $q$ (B);  in the case of the $\beta$--hairpin using the dRMSD (C) and $q$ (D).  Spurious drift amounts to $\sim 10^{-1}$ and thus does not affect the comparison. All energies are in arbitrary units. Dash--dotted lines indicate the diagonal of the plot.}
\label{fig:correlations}
\end{figure*}

The equilibrium free energy $F$ as a function of  dRMSD and $q$  for the $\alpha$--helix and the $\beta$--hairpin were obtained from replica--exchange simulations \cite{Sugita:1999} with a weighted--histogram algorithm \cite{Sun:2014}. They are displayed in the bottom panels of Figs. \ref{fig:helix_plots}  and \ref{fig:hairpin_plots}, respectively.  In Fig. \ref{fig:correlations} we compared the effective mean force $\gamma D^{(1)}=TD^{(1)}/D^{(2)}$ with the derivative $-dF/dY$ obtained from the numerical differentiation of the values displayed in Figs. \ref{fig:helix_plots} and \ref{fig:hairpin_plots}. 
In spite of the large standard deviation which affects the distributions of 
$D^{(1)}$ and  $D^{(2)}$ (cf.  Figs. \ref{fig:helix_plots} and \ref{fig:hairpin_plots}), 
the two quantities are comparable, relative errors being in the worst case of the order of 50\% of the force. The dRMSD performs worse than $q$. 
Moreover, in three cases the two are fairly correlated (the correlation coefficients being $r>0.89$), while only for the dRMSD of the $\beta$--hairpin the correlation seems poor.

It is worth to point out that the two quantities compared in Fig. \ref{fig:correlations} have a completely different origin. The effective force $TD^{(1)}/D^{(2)}$ is calculated point--wise from dynamical simulations, while the free energy is obtained from an equilibrium sampling. Moreover, the agreement occurs in spite of the large width of the distribution of drift and diffusion coefficients. 

\begin{figure*}
\includegraphics[width=0.75\textwidth]{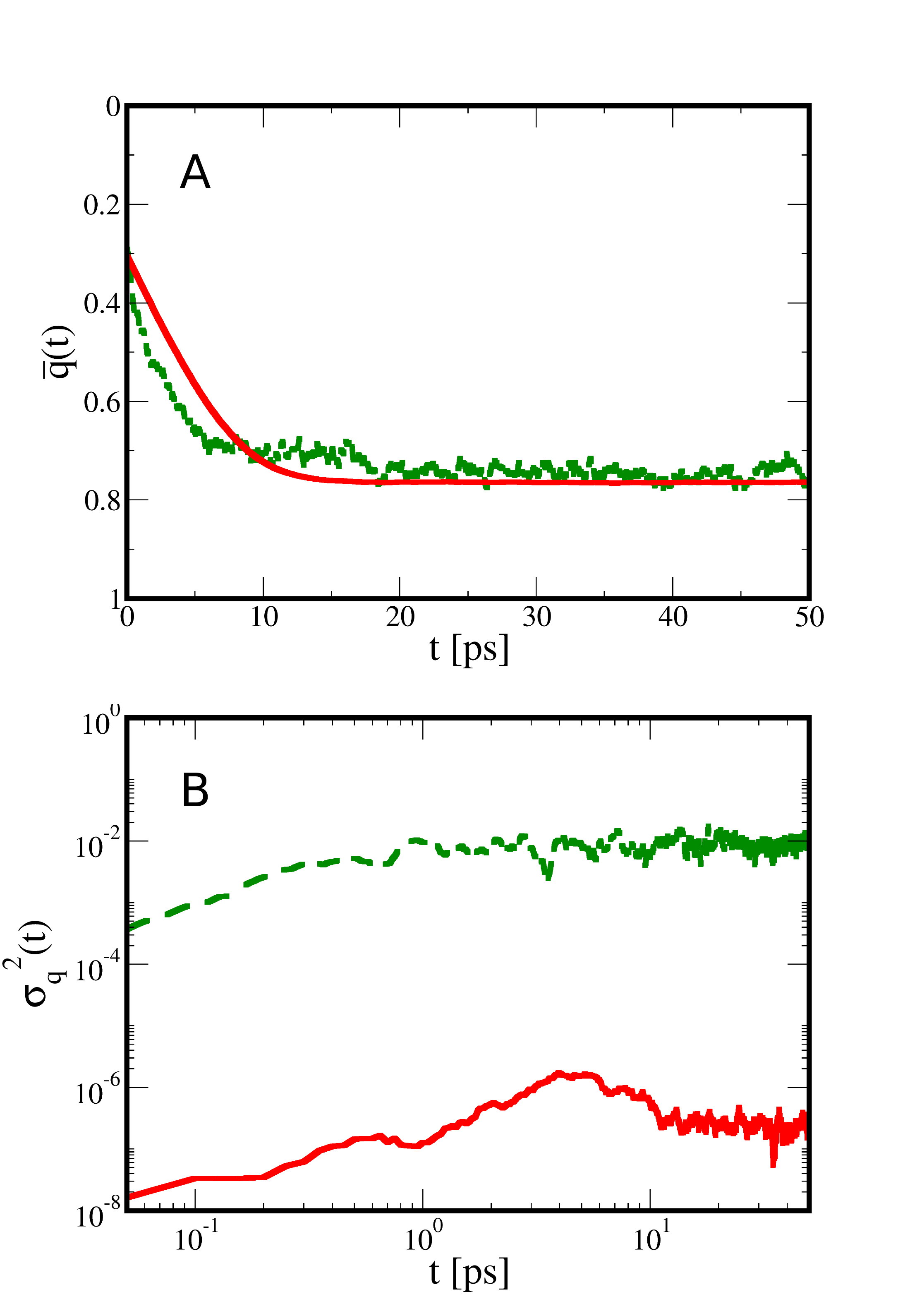}
\caption{
The average fraction of native contacts $\bar{q}$ (A) and its variance (B), 
as functions of the simulation time. 
Data displayed with green (dashed) lines are  taken from full--atom simulations of the $\beta$-hairpin, while 
 red (solid) lines are generated via the resolution of a monodimensional langevin equation.
The relaxing times $\tau$ of $\bar{q}$ are 
$\tau_{\beta}=3.6$ ps for the full--atomistic simulations and
$\tau_{1d}=4.8$ ps for the monodimensional trajectories.}
\label{fig:kinetics}
\end{figure*}

As a further test for the drift and diffusion coefficient we found, we compared the dynamics simulated by Eq.  (\ref{eq:red-langevin}) with the projection of the microscopic dynamics on the CV. This comparison is carried out for the case of the variable $q$ in the $\beta$--hairpin, which gave the best results in the determination of the effective energy (cf. Fig. \ref{fig:correlations}D). In Fig. \ref{fig:kinetics} it is displayed the average and the standard deviation of $q$ as a function of time, over 100 simulations starting from the same unfolded conformation. Although the overall folding time seems comparable in the two kind of simulations (cf. Fig. \ref{fig:kinetics}A), the detailed dynamics is different. In particular, the microscopic dynamics displays a slightly faster initial folding, followed by a further event around 18 ps, which is not reported in the effective Langevin dynamics. However, it is the run--to--run variability to be completely different in the two cases (cf. Fig. \ref{fig:kinetics}B), being orders of magnitude smaller in the microscopic dynamics.

The dynamical properties thus seem much more sensitive to the imperfections in the reaction coordinate than to the equilibrium properties.

\section{Discussion}

The results obtained in two one--dimensional test systems indicate that it is possible to back--calculate the diffusion coefficient with great precision (i.e., with an error lower than 1\%) by direct application of its finite--difference definition (i.e., in Eq. (\ref{eq:km2})) if the drift coefficient is moderately small, that is if the drift does not induce a deterministic displacement within the time $\Delta t$ used in Eq. (\ref{eq:km2}). 

On the other hand, the calculation of the drift coefficient directly by definition would require an enormous ($\gg 10^5$) number of replicated simulations for each point of conformational space to obtain a stable result, and it is thus impractical. The strategy we developed based on an iterative solution of the finite--differences equations of motion allow one to back--calculate the drift coefficient with an error lower than 10\% in the test cases, and to back--calculate correctly the diffusion coefficient also if the system undergoes strong drifts, up to those caused by nN forces, much stronger than typical biological forces. 

In simulations carried out with simple models of alpha--helix and beta--hairpin, the distribution of drift coefficients for different point of conformational space associated with the same value of the fraction of native contacts $q$ has a spread which is of the order of half its average. This suggests that the CV $q$ defines a drift coefficient, although with a non--negligible error bar.  The same is not true for the dRMSD. In this case the drift coefficient is defined only in terms of its order of magnitude, the width of its distribution for fixed value of dRMSD ranging from three to ten times the mean. 

Anyway, for both variables the mean drift matches that obtained as derivative of the free energy, calculated independently from a conformational sampling at equilibrium. This suggests that it is possible to build a one--dimensional approximated model of these peptides, in which the equilibrium and the dynamical properties are consistent with each other. Moreover, one could also exploit these results to calculate the equilibrium free energy of a system from short dynamical simulations.

The diffusion coefficient is defined better than the drift  for both the dRMSD and the $q$ , the width of the associated distribution being in both cases at worst half of the mean, and usually lower.  This fact gives a sound basis to the calculations reported in refs. \cite{Hummer:2005eb,Sriraman:2005jf,Best:2006gj,Best:2010cxa} for the diffusion coefficient, and make it possible to exploit strategies in which the diffusion coefficient is artificially biased to enhance conformational sampling.

The large widths of the distributions of $D^{(1)}$ and $D^{(2)}$ are not really unexpected, since to a given value of $q$ or dRMSD correspond conformations which can be conformationally very different, and thus can display different energetic properties. Interestingly this is true 
for native--like conformations as well ($q=0.90$ or dRMSD$=0.20$ nm), which are supposedly more homogeneous from the conformational point of view. Most likely, the steep dependence of the potential function which is used in the present force field (i.e., containing terms like $1/r^6$ or $1/r^{12}$) and which reflect the true interaction between the atoms of the system, plays an important role in defining the width of the observed distributions.

Although the agreement between the calculated mean drift coefficient and the equilibrium free energy of the peptides is reasonably good, the detailed dynamics is quite different in the dimensional--reduced model, especially in terms of run--to--run fluctuations of the CV. This suggests that the dynamical properties are more sensitive to approximations than the equilibrium properties. This asymmetry was already observed for their respective dependence on the force field \cite{Piana:2011}.

\section{Conclusions}

The dimensional--reduction approach can be very useful to analyze complex biomolecular 
data and also to bias MD simulations in order to decrease their computational cost. 
The dynamics of a system in a reduced dimensional space can be as complicated 
as that in the original, full--dimensional space if the reduced coordinate $Y$ 
is not chosen carefully. If $Y$ describes the slowest motion of the system, its dynamics 
is controlled by a simple Langevin equation. However this condition is difficult 
to check directly in a system as complex as a biopolymer. 
Here, we employed a different approach. We developed a method to calculate 
the drift and diffusion coefficient in the neighborhood of a microscopic point 
of the full--dimensional conformational space. This method is based on a finite--difference
version of the Langevin equation and an iterative evolution of the dynamics of $Y$ under the
approximation of locally linear force and locally constant diffusion coefficient.
This allowed us to calculate the drift and diffusion coefficients for an ensemble of such points 
corresponding to the same value of the reduced coordinate $Y$. 
We showed that, as already reported, the coefficients calculated for the length--invariant 
collective coordinates (like the contact function $q$) display a weaker dependence 
on the microscopic point than Euclidean distances (like the dRMSD), 
but anyway they are not negligible. 
Nonetheless, the average drift coefficients are compatible with the equilibrium properties of the system. On the other hand, the dynamical properties are more sensitive to the lack of ideality of the reaction coordinate.

\begin{acknowledgments}
CC is  supported by the Technische Universit{\"a}t M{\"u}nchen -- Institute for Advanced Study, funded by the German Excellence Initiative and the European Union Seventh Framework Programme under grant agreement n. 291763.
\end{acknowledgments}


\newpage

\end{document}